\documentclass[twocolumn,aps,prl,superscriptaddress]{revtex4-2}
\usepackage{color}
\usepackage{mathrsfs}
\usepackage{bm}
\usepackage{amsmath}
\usepackage{amssymb}
\usepackage{graphicx}
\usepackage{hyperref}
\usepackage{color}
\usepackage{siunitx}
\usepackage{braket}
\usepackage{tikz}
\usepackage{array} 
\usepackage{booktabs}
\allowdisplaybreaks[3]

\makeatletter
\hypersetup{hypertex=true, colorlinks=true, linkcolor=blue, anchorcolor=blue,citecolor=blue}

\usepackage{dcolumn}
\usepackage{bm}
\usepackage{mathrsfs}
\usepackage{amsfonts}
\usepackage{xcolor}
\usepackage{epstopdf}

\begin{document}
\title{Higher-Order Topological Superconductivity and Electrically Tunable Majorana Corner Modes in Monolayer MnXPb$_2$ (X=Se, Te)-Pb Heterostructure} 
\author{Yongting Shi}
\affiliation{Institute of Applied Physics and Computational Mathematics, Beijing 100088, China}
\affiliation{Anhui Provincial Key Laboratory of Low-Energy Quantum Materials and
		Devices, High Magnetic Field Laboratory, HFIPS, Chinese Academy
		of Sciences, Hefei, Anhui 230031, China}
\author{Qing Wang}
\affiliation{Anhui Provincial Key Laboratory of Low-Energy Quantum Materials and
		Devices, High Magnetic Field Laboratory, HFIPS, Chinese Academy
		of Sciences, Hefei, Anhui 230031, China}
\affiliation{Science Island Branch of Graduate School, University of Science and
		Technology of China, Hefei, Anhui 230026, China}

\author{Zhen-Guo Fu} 
\email{fu\_zhenguo@iapcm.ac.cn}
\affiliation{Institute of Applied Physics and Computational Mathematics, Beijing 100088, China}
\author{Ping Zhang}
\email{zhang\_ping@iapcm.ac.cn}
\affiliation{Institute of Applied Physics and Computational Mathematics, Beijing 100088, China}
\affiliation{National Key Laboratory of Computational Physics, Beijing 100088, China}
\affiliation{School of Physics and Physical Engineering, Qufu Normal University, Qufu 273165, China}
\author{Ning Hao}
\email{haon@hmfl.ac.cn}
\affiliation{Anhui Provincial Key Laboratory of Low-Energy Quantum Materials and
		Devices, High Magnetic Field Laboratory, HFIPS, Chinese Academy
		of Sciences, Hefei, Anhui 230031, China}
		
		
\begin{abstract} 
Higher-order topological superconductors host Majorana zero modes localized at corners or hinges, providing a promising route toward scalable and controllable Majorana networks without vortices or magnetic flux. Here we propose a symmetry-enforced higher-order topological superconductivity based on antiferromagnetic topological insulators, specifically realized in MnXPb$_2$ (X = Se, Te)-Pb heterostructure. We show that the intrinsic boundary dichotomy—gapless Dirac states protected by an effective time-reversal symmetry on antiferromagnetic edges and magnetic gaps on ferromagnetic edges—naturally generates Majorana corner modes as mass domain walls.
Superconducting proximity converts the antiferromagnetic edges into one-dimensional topological superconductors, and the intersections between superconducting and magnetic edges bind Majorana zero modes as mass domain walls.
Combining first-principles calculations with a calibrated effective boundary theory, we demonstrate robust corner localization and purely electrical control of Majorana fusion and braiding in a triangular geometry. Our results establish MnXPb$_2$ as experimentally promising platform for electrically programmable Majorana networks in two dimensions.
	\end{abstract}
	\maketitle
	
	\textit{Introduction.}--- The realization of topological superconductivity and Majorana zero modes (MZMs) is a central objective of condensed matter physics due to their non-Abelian exchange statistics and potential applications in fault-tolerant quantum computation\cite{TS1,TS2,TS3,TS4,TS5,TS6,TS7,TS8, TS10, TS9}. Over the past decade, Majorana modes have been intensively explored in a variety of platforms, including topological insulator-superconductor heterpstructure\cite{TS35, PhysRevLett.114.017001, PhysRevLett.116.257003}, semiconductor–superconductor nanowires under Zeeman fields\cite{sau2010generic}, magnetic-atom chains on superconducting substrates\cite{nadj2014observation}, and iron-based superconductors with topological surface states\cite{hao2014topological, wang2015topological, wu2016topological, hao2019topological, wu2015cafeas, xu2016topological}. While these systems have demonstrated key signatures of Majorana physics, most existing proposals rely on vortices, Zeeman fields, or complex junction networks, which present significant challenges for scalability, tunability, and device integration.
	
A particularly appealing direction is offered by higher-order topological superconductors (HOTSCs)\cite{PhysRevB.98.165144, yan2019higher, zhang2020higher, khalaf2018higher, scammell2022intrinsic, ahn2020higher, li2024realizing, wong2023higher, pan2019lattice}. Unlike conventional topological superconductors that host gapless modes along their boundaries, HOTSCs support zero-dimensional MZMs localized at corners in two dimensions or hinges in three dimensions\cite{PhysRevLett.124.227001, PhysRevLett.123.167001, TS82,TS83, TS88}. These modes emerge from mass domain walls formed by competing edge gaps and provide a natural route toward network architectures in which fusion and braiding operations can be implemented through geometric design and local parameter control, without relying on vortices\cite{PhysRevLett.114.017001, PhysRevLett.116.257003, wang2018evidence, li2022ordered, liu2024signatures} or nanowire endpoints\cite{karzig2017scalable, TS84,TS85,TS86,TS87, song2022phase}. Several theoretical routes to HOTSCs have been proposed, including in-plane Zeeman-field engineering in s-wave heterostructures\cite{ PhysRevB.108.184517, volpez2019, PhysRevLett.124.227001}, and iron-based superconductors with intrinsic topological surface states\cite{PhysRevLett.123.167001, wu2020boundary}. Despite their conceptual significance, these approaches remain constrained by material specificity and the need for external magnetic fields, which are often detrimental to superconductivity\cite{machida2019zero, oyler2009chemical}. 

A central open challenge is therefore to identify a two-dimensional Majorana platform that is simultaneously robust, electrically controllable, and compatible with conventional superconducting proximity effects, without relying on Zeeman fields or vortices. 

In this work, we introduce a distinct route to HOTS rooted in the symmetry and boundary structure of antiferromagnetic topological insulators (AFMTIs).
We demonstrate that AFMTIs possess an intrinsic boundary dichotomy: antiferromagnetic edges support gapless Dirac modes protected by an effective time-reversal symmetry, whereas ferromagnetic edges are generically gapped by magnetic mass terms. When superconducting proximity is introduced, the antiferromagnetic edges are converted into one-dimensional topological superconductors, while the ferromagnetic edges remain insulating. Their intersections thus form symmetry-enforced mass domain walls that bind Majorana corner modes (MCMs). This mechanism does not rely on Zeeman-field engineering or momentum-space self-proximity and constitutes a general route to HOTSCs in two dimensions.

As a concrete realization of this symmetry-enforced mechanism, we identify monolayer MnXPb$_2$ (X = Se, Te) as an experimentally accessible AFMTI platform proximitized by a conventional Pb superconductor. Using first-principles calculations, topological invariants, and an effective theory calibrated to realistic energy scales, we demonstrate robust MCMs and purely electrical control of their fusion and braiding in a triangular geometry. Our results establish AFMTIs as a generic materials platform for electrically programmable Majorana networks.

\begin{figure}[htp]
\begin{center}
	\includegraphics[width=1.0\linewidth]{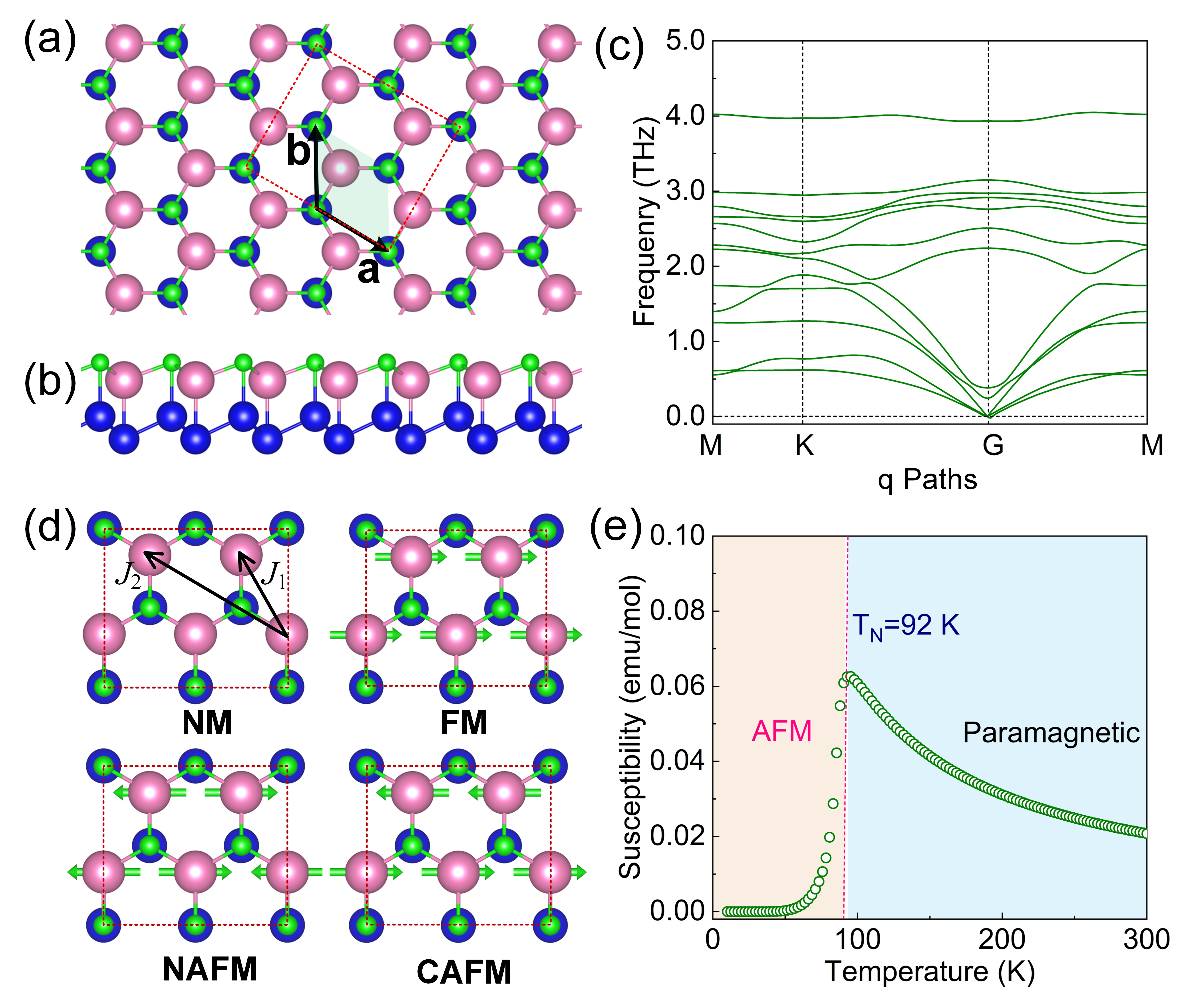}
\end{center}
	\caption{Crystal structure, magnetic ground state, and stability of monolayer MnXPb$_2$ (X = Se, Te).
  (a) Top and (b) side views of the trigonal crystal structure (space group P3m1), where all Mn atoms lie in a single plane and bond to three neighboring chalcogen atoms. Pink, green and
blue balls denote Mn, Te and Pb atoms, respectively.
  (c) Phonon dispersion without imaginary modes, confirming dynamical stability.
  (d) Collinear in-plane antiferromagnetic ground state identified from total-energy comparisons among nonmagnetic, ferromagnetic, N\'eel antiferromagnetic, and collinear antiferromagnetic configurations. $J_1$, $J_2$ denote exchange coupling between first- and second-nearest neighbor Mn atoms.
  (e) Temperature-dependent magnetic susceptibility fitted by a Curie--Weiss form, yielding a N\'eel temperature $T_N \approx 92$ K.}
	\label{fig-1}
\end{figure}

\textit{Platform and Higher-Order Topology.}--- Monolayer MnXPb$_2$ crystallizes in a trigonal structure with space group  \emph{P}$\overline{3}$\emph{m}1 (No. 156). As shown in Fig. \ref{fig-1}(a) and (b), all Mn atoms lie in a single plane and are bonded to three neighboring chalcogen atoms (X = Se or Te), while Pb atoms occupy two adjacent planes, forming a layered geometry reminiscent of AA-stacked bilayer silicon. Structural optimization yields lattice constants 
$a=b=4.72$ \AA, with Mn–X and Pb–Pb bond lengths of 2.90 \AA~and 3.01 \AA (for X = Te), respectively, both shorter than the Mn–Pb bond length of 3.48 \AA. This hierarchy indicates strong intralayer coupling and weak interlayer hybridization, a favorable condition for van der Waals heterostructures.

Phonon dispersion calculations exhibit no imaginary frequencies across the Brillouin zone [Fig. \ref{fig-1}(c)], confirming dynamical stability. We next examine the magnetic ground state by comparing four representative configurations: nonmagnetic (NM), ferromagnetic (FM), Néel antiferromagnetic (NAFM), and collinear antiferromagnetic (CAFM), defined in a 
$(2\bm{a},2\bm{b}+\bm{a})$ supercell (Fig. \ref{fig-1}(d)). Total-energy comparisons identify the CAFM configuration as the ground state. Magnetic anisotropy calculations further show that in-plane spin orientation is favored by 2.62 meV per four Mn atoms relative to out-of-plane orientation, consistent with related Mn–Te compounds.

The CAFM order can be captured by a Heisenberg model with nearest-neighbor ($J_1$) and next-nearest-neighbor ($J_2$) couplings, yielding energies $E_{FM}=-(6J_1+6J_2)S^2$, $E_{NAFM}=(-2J_1+2J_2)S^2$, and $E_{CAFM}=(-2J_1-2J_2)S^2$ with $S=5$ per Mn atom. Fitting to the first-principles energies gives $J_1=-0.67$ meV and $J_2=-2.54\times10^{-6}$ meV, confirming the CAFM ground state. Temperature-dependent susceptibility yields a Néel temperature $T_N=92$ K [Fig. \ref{fig-1}(e)], providing a robust magnetic energy scale.  
 
\begin{figure}[htp]
\begin{center}
	\includegraphics[width=1.0\linewidth]{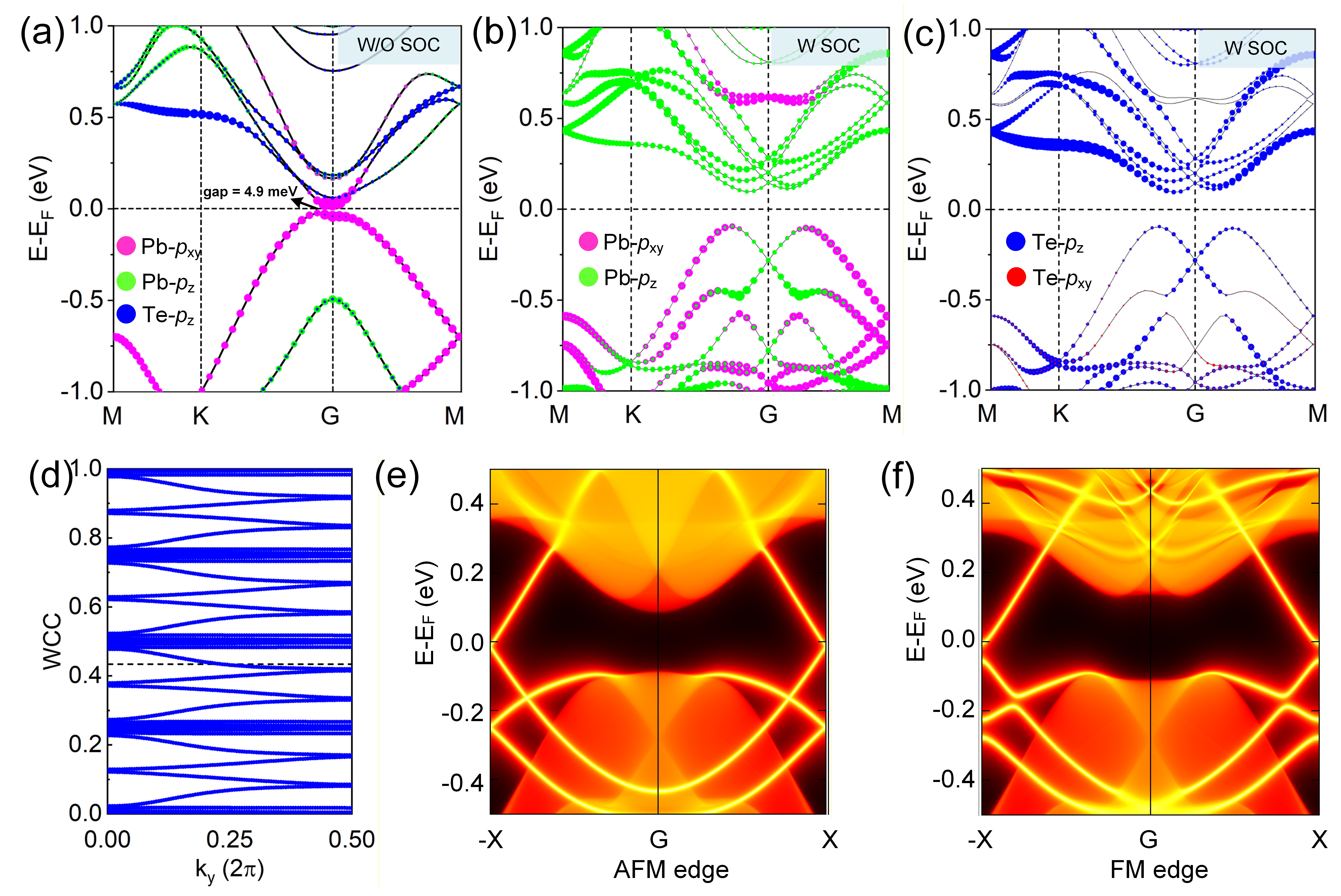}
\end{center}
	\caption{Topological electronic structure and boundary dichotomy in monolayer MnTePb$_2$.
  (a) Band structure without spin--orbit coupling in the collinear antiferromagnetic state.
  (b) Band structure with spin--orbit coupling, showing an indirect gap of about 190 meV.
  (c) Orbital-resolved band character near the Fermi level, indicating SOC-driven band inversion.
  (d) Evolution of hybrid Wannier charge centers, yielding a nontrivial $Z_2=1$ invariant protected by an effective time-reversal symmetry.
  (e) Edge spectrum for an antiferromagnetic boundary, hosting a Dirac crossing at the X point.
  (f) Edge spectrum for a ferromagnetic boundary, exhibiting a full gap.}
	\label{fig-2}
\end{figure}

We next analyze the electronic structure of freestanding MnXPb$_2$ in the CAFM state. Without spin–orbit coupling (SOC), the system is an antiferromagnetic narrow-gap semiconductor [Fig. \ref{fig-2}(a)]. Including SOC opens an indirect gap of approximately 190 meV [Figs. \ref{fig-2}(b,c)] and drives an orbital-weight redistribution near the $\Gamma$ point, signaling a topological phase. Because the CAFM order breaks conventional time-reversal symmetry, the topology is protected by an effective time-reversal operation that combines time reversal with a half-unit-cell translation.

We compute the $Z_2$ invariant using the evolution of one-dimensional hybrid Wannier centers and obtain 
$Z_2=1$ [Fig. \ref{fig-2}(d)], confirming a topological insulating phase. The phase remains robust under biaxial strain from -6\% to 6\% (see Supplemental Material). Edge-state spectra reveal a boundary dichotomy: antiferromagnetic edges host gapless Dirac modes at the X point, whereas ferromagnetic edges are gapped [Figs. \ref{fig-2}(e,f)]. This dichotomy is the central ingredient for higher-order topology.

From a symmetry perspective, the CAFM order admits an effective time-reversal symmetry $\Theta$$_M$ = $\Theta$$e^{i(k_{x}-\pi)/2}$, where $\Theta$ denotes time reversal and the exponential encodes a half-translation along the edge. On antiferromagnetic edges, $k_x$ is a good quantum number and $\Theta_{M}^{2}=-1$ at the $X$ point, enforcing Kramers degeneracy and protecting the Dirac crossing. On ferromagnetic edges, 
$k_x$ is not a good quantum number and the antisymmetry condition fails, allowing a mass gap. 
 
\begin{figure}[htp]
\begin{center}
	\includegraphics[width=1.0\linewidth]{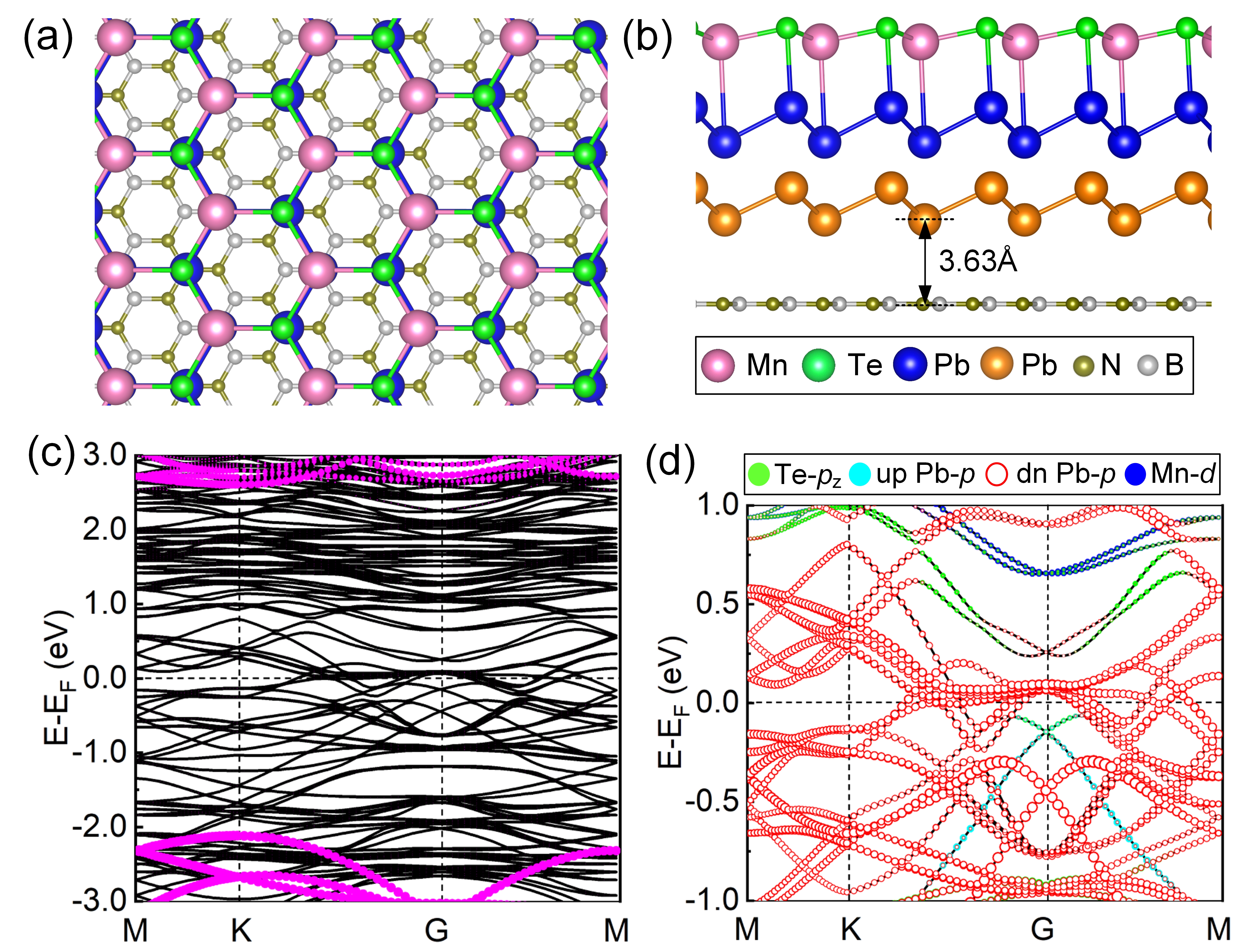}
	\end{center}
	\caption{MnTePb$_2$--Pb heterostructure and proximity regime.
  (a) Top and (b) side views of the relaxed stable MnTePb$_2$--Pb--BN heterostructure constructed from commensurate supercells.
  (c) Band structure of the heterostructure with orbital projections, showing weak normal-state hybridization between MnTePb$_2$ and Pb.
  (d) Enlarged view near the Fermi level, illustrating that MnTePb$_2$ boundary states remain well separated from Pb bands, a favorable condition for proximity-induced superconductivity.}
\label{fig-3}
\end{figure}

\textit{Proximity Effect and Interface Feasibility}--- The antiferromagnetic boundary of MnTePb$_2$ hosts Dirac-type edge stat [Fig. \ref{fig-2}(e)], which can form a one-dimensional topological superconductor under proximity-induced superconductivity, and corners formed between superconducting and ferromagnetic edges act as zero-dimensional mass domain walls and bind MCMs. To realize superconducting proximity, we construct a MnXPb$_2$-Pb heterostructure on a BN substrate using commensurate supercells: 4×4 BN, 2×2 Pb, and 2×2 MnTePb$_2$ supercell. After van der Waals relaxation, the interface is stable with a lattice mismatch of $-$3.4\%, and MnTePb$_2$ preserves its intrinsic geometry [Fig. \ref{fig-3}(a, b)]. BN serves as an inert buffer that minimizes substrate-induced disorder and facilitates epitaxial growth.

The projected band structure of the heterostructure is shown in Fig. \ref{fig-3}(c). MnTePb$_2$ bands remain essentially unchanged, while Pb states cross the Fermi level and exhibit weak coupling to the MnTePb$_2$ layer [Fig. \ref{fig-3}(d)]. This separation indicates weak normal-state hybridization, a favorable condition for proximity-induced superconductivity below the Pb transition temperature. In this regime, Cooper pairs leak into the MnTePb$_2$ boundary states, opening a superconducting gap while preserving their topological character.

The large bulk gap of MnTePb$_2$ further suppresses bulk–edge mixing, ensuring that the induced one-dimensional topological superconducting channels are spatially confined along antiferromagnetic edges. Together, the stable interface, weak normal-state hybridization, and large bulk gap place the proposed platform squarely within the experimentally accessible parameter window of van der Waals heterostructures and thin-film superconductors.

\begin{figure}[htp]
\begin{center}
	\includegraphics[width=1.0\linewidth]{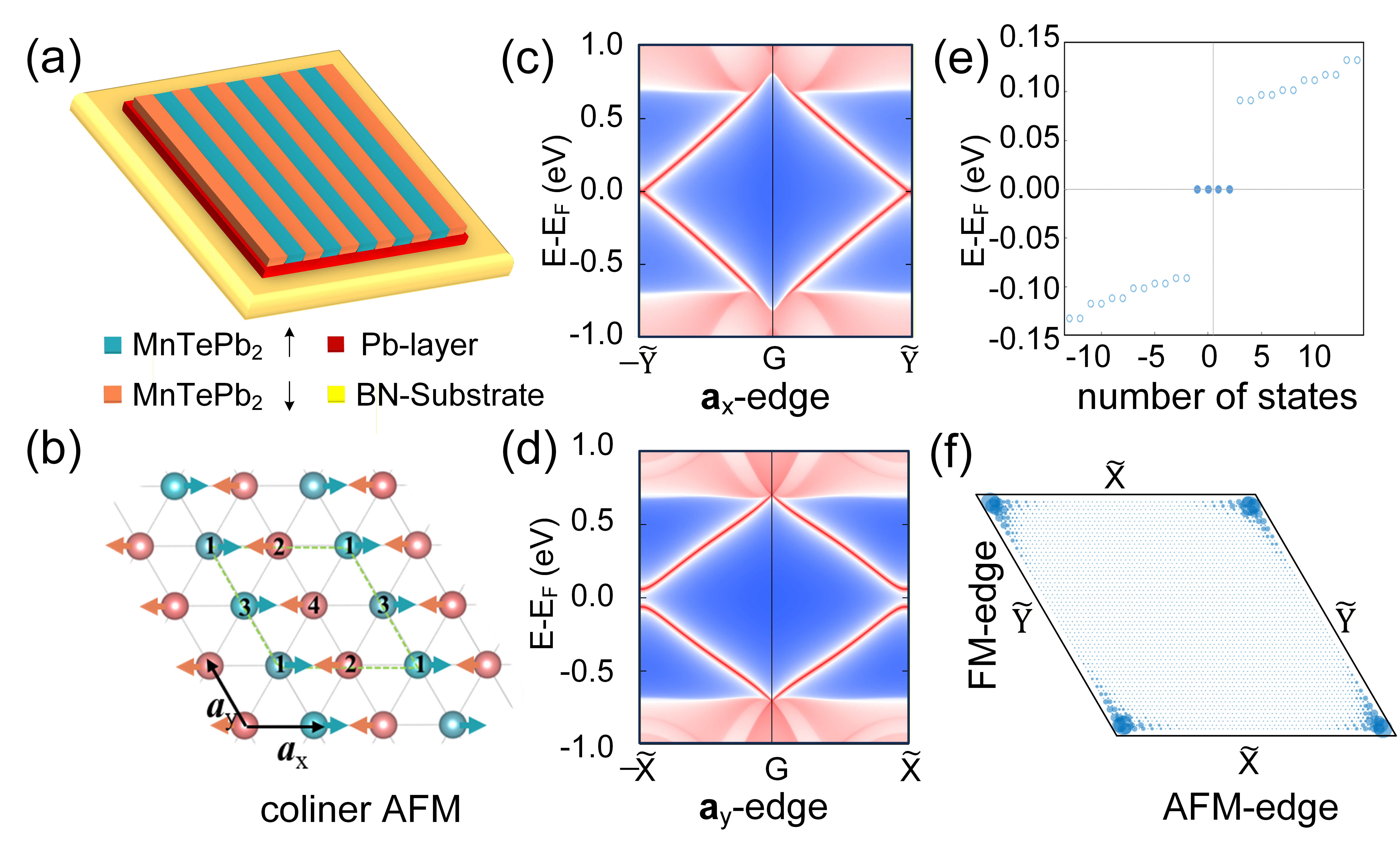}
	\end{center}
	\caption{Boundary model and Majorana corner modes in the higher-order topological phase.
  (a) Schematic of the triangular lattice and collinear antiferromagnetic unit cell.
  (b) Folded Brillouin zone used to construct the effective boundary model.
  (c) Spectrum of an antiferromagnetic edge without superconductivity, hosting a Dirac mode protected by an effective time-reversal symmetry.
  (d) Spectrum of a ferromagnetic edge, showing a magnetic mass gap.
  (e) Low-energy spectrum of a finite cluster in the presence of superconducting proximity, revealing four zero-energy states inside the edge gap.
  (f) Spatial probability density of the four zero modes, demonstrating exponential localization at the corners. In (c)-(d), the parameters ($A$, $B$, $m$, $M$, $\mu$)=(1, 1, 3, 0.1, 0). In (e)-(f), the parameters ($A$, $B$, $m$, $M$, $\mu$, $\Delta$ )=(1, 1, 3, 0.2, 0, 0.1)}
\label{fig-4}
\end{figure}

 \textit{Effective Model and Majorana Corner Modes}
To analyze the proximitized boundaries, we derive an effective model in the unfolded Brillouin zone under CAFM order [Fig. \ref{fig-4}(a)].
The projection relation between the unfold and fold Brillouin zone is discussed in details in SMs. The resulting effective Hamiltonian kernal capturing the SOC-induced band inversion, the antiferromagnetic exchange field, and proximity pairing, can be reduced into a 8×8 matrix form (See SMs for details), 
 \begin{equation}
\begin{aligned}
	H(\bm{k})=[ & (m-6 B)+2 B\sum_{i=1}^{3} \cos \bm{k} \cdotp{\bm{d_i}}] \tau_{\mathrm{z}} \sigma_{z} \\
	& +A\left(\sum_{i=1}^{3}\sin \bm{k} \cdotp{\bm{d_i}}\left(\cos \theta_{i} \sigma_{x} s_{z}-\sin \theta_{i} \tau_{z} \sigma_{y}\right)\right) \\
	& -\mu \tau_{z}+\Delta \tau_{y} s_{y}+\frac{M}{\sqrt{1+\alpha^{2}}} \tau_{z} (\sigma_{0}+\alpha \sigma_{z}) s_{x}.
\end{aligned}
 \label{func_1}
\end{equation}
Here, the three nearest-neighbor vectors of the triangular lattice are: $\bm{{d}_{1}}=\bm{{a}_{1}}=(1,0)$, $\bm{{d}_{2}}=\bm{{a}_{2}}=\left(-\frac{1}{2}, \frac{\sqrt{3}}{2}\right)$, $\bm{{d}_{3}}=\bm{{a}_{1}}+\bm{{a}_{2}}=\left(\frac{1}{2}, \frac{\sqrt{3}}{2}\right)$. Their direction angles $\theta_i$ (where i = 1, 2, 3) are 0, 2$\pi$/3, and $\pi$/3, respectively. Three types of Pauli matrices $\bm{\sigma}$, $\bm{\tau}$, $\bm{s}$ span the orbital, particle-hole and spin sub-space, respectively. The effective model in Eq. (\ref{func_1}) is quantitatively anchored to first-principles calculations of MnXPb$_2$. The Dirac velocity and mass parameters $A$, $B$, $m$ are obtained by fitting the low-energy DFT band dispersions in the vicinity of the Fermi level, reproducing the SOC-induced band inversion and the bulk gap\cite{TS75}. The exchange field $M$ is estimated from the collinear antiferromagnetic ground state and the transition temperature, providing an energy scale well above the induced superconducting gap. The proximity-induced pairing amplitude $\Delta$ is taken from experimentally measured gaps of few-layer Pb films, serving as an upper bound for the induced edge gap. $\alpha$ is used to consider the possible differences in the $g$ factors of $p$-electrons and $d$-electrons. Since there is almost no physical impact, we also take $\alpha$ = 0 for simiplicity. The energy scales of the model parameters can bee stimated for MnTePb$_2$. $A$, $B$ and $m$ take the values of $-95.35$ meV, 32 meV and 266 meV, respectively. We verify that the higher-order topological phase and the corner modes remain robust over a broad range of parameters around these experimentally relevant values, indicating that the effect does not rely on fine tuning.

Without superconductivity, the model reproduces the boundary dichotomy observed in first-principles calculations: antiferromagnetic edges host gapless Dirac modes protected by the effective time-reversal symmetry, whereas ferromagnetic edges are gapped [Figs. \ref{fig-4}(c,d)]. With proximity pairing, the antiferromagnetic edge becomes a one-dimensional topological superconductor. On the ferromagnetic edge, the magnetic mass competes with the superconducting mass. When the magnetic mass dominates, the ferromagnetic edge remains gapped in the normal state, and the intersection between a superconducting antiferromagnetic edge and a magnetic ferromagnetic edge forms a zero-dimensional domain wall. This domain wall binds a single Majorana zero mode, realizing a second-order topological superconducting phase.

We confirm this mechanism by diagonalizing the model on a finite  40$\widetilde{a}_x$ $\times$ 40$\widetilde{a}_y$ cluster with open boundaries. In the topological regime, four zero-energy states appear within the edge gap [Fig. \ref{fig-4}(e)]. Their combined spatial profile exhibits exponential localization at the four corners [Fig. \ref{fig-4}(f)], with negligible weight in the bulk and along the edges. The localization length is set by the inverse of the smaller of the superconducting and magnetic masses, ensuring well-separated zero modes for realistic parameters.

\textit{Electrical Control and Braiding.}---The low-energy $\bm{k}\cdotp \bm{p}$ Hamiltonian from the tight-binding Hamiltonian in Eq. (\ref{func_1}) has the same form as that in Ref. \cite{TS82}, from which, we can obtain the similar motion rule for the MCMs on a single triangular island under the electrical control of $\mu$.  It should be noted that the critical angle of our isosceles obtuse triangle is 60 degrees, which is different from the 45 degrees of the isosceles right triangle in Ref. \cite{TS82}. Therefore, the motion of the MCMs can be controlled by tuning the chemical potential 
$\mu$ relative to its critical value $\mu_c = (3M^2/4 - \Delta^2)^{1/2}$. Such controlling has been confirmed through numerical calculations shown in SMs. Using experimental parameters, ARPES on monolayer Pb gives $\Delta \approx$ 0.8meV \cite{TS89}. Néel temperatures of 92 K of MnTePb$_2$ corresponds to M $\approx$ 7.92 meV—well above the SC gap—ensuring the topological condition is met for small $|\mu|$. 

\begin{figure}[htp]
\begin{center}
	\includegraphics[width=1.0\linewidth]{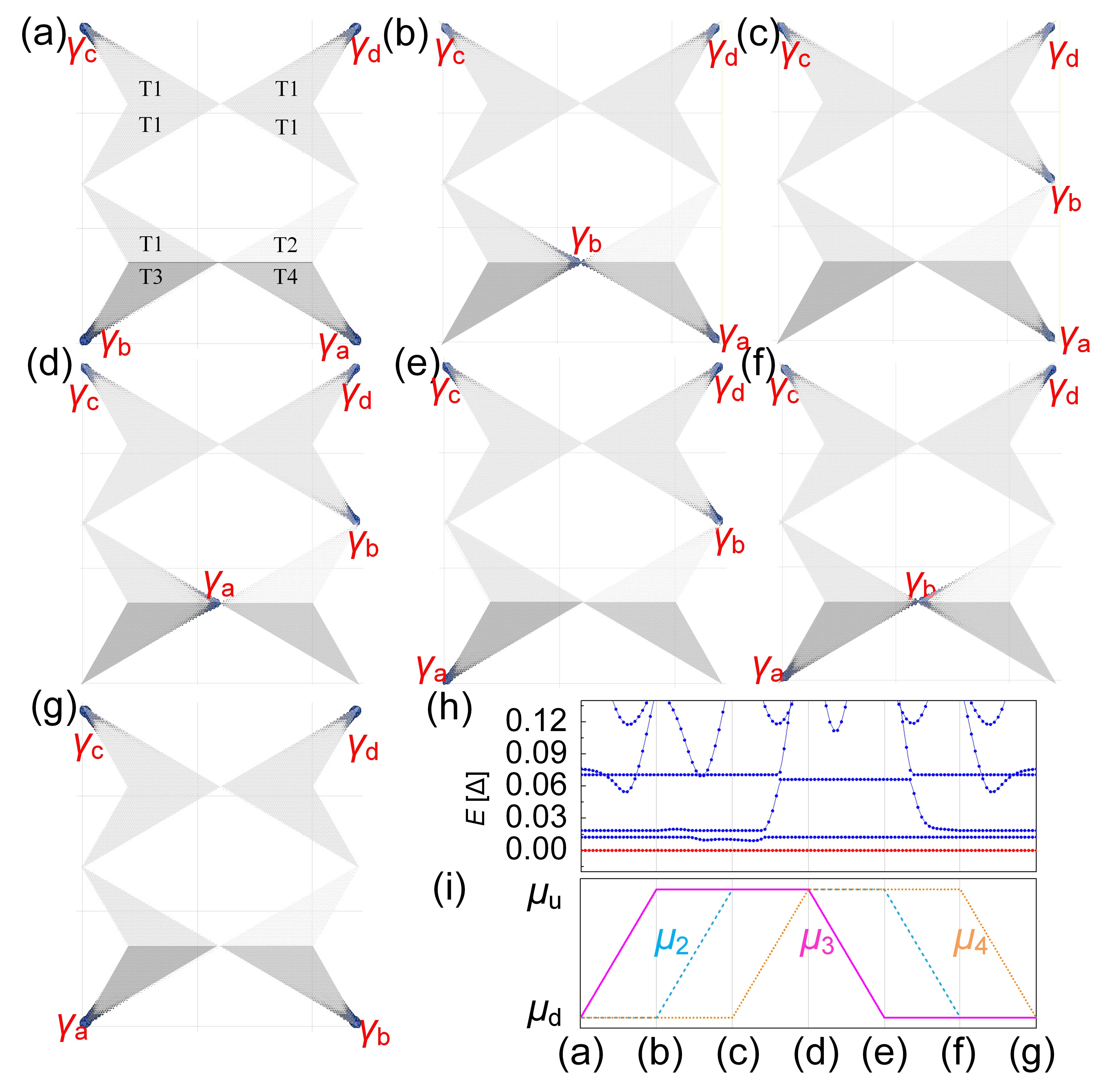}
	\end{center}
	\caption{ Electrical control and adiabatic braiding of Majorana corner modes.
  (a--g) Real-space evolution of Majorana probability densities during an adiabatic braiding cycle implemented by tuning local chemical potentials on three selected triangular islands.
  The remaining islands are kept fixed.
  (h) Energy spectrum along the braiding path, showing that the four Majorana modes remain pinned at zero energy and separated from excited states by a finite gap.
  (i) Time sequence of the chemical-potential protocol used in the braiding operation.
During the protocol, $\mu_1$ = $\mu_d$ is fixed. $\mu_2$, $\mu_3$, and $\mu_4$ are varied in time according to (i). The parameters are $\mu_u$ = 0.2$m$, $\mu_d$ = 0.05$m$, M = 0.4$m$, $\Delta$ = 0.25$m$, and A =  B = 0.5$m$ = 3; the short-side length of the IRTIs is L = 40a}
\label{fig-5}
\end{figure}

With two islands at finite phase difference $\delta\Phi$, inter-island fusion of two overlapped MCMs become possible and fusion strength $F$$_{\varphi_i:\varphi_j'}$ follow the same dependence in Refs. \cite{TS82,TS83}. Such phase-controlled couplings enable multi-MCM braidings. Usually, braiding in typical 2D platforms requires a complex network \cite{TS84,TS85,TS86,TS87,TS88}. Here, we design a simple eight-isosceles-obtuse-triangle-islands (IOTIs) prototype device, as shown in Fig. \ref{fig-5}(a),  to realize the minimum braiding in the space spanned by four MCMs. 
The IOTIs are labeld by T$_j$ (where j $\in$ \{1, $\cdots$, 4\}) with local chemical potential $\mu_j$ and and superconducting phase $\Phi_j$, respectively. For the numerical simulation shown in Fig. \ref{fig-5}, we considered the cases where $\Phi_3$ = $\Phi_4$ = $\pi$/2 and $\Phi_j$ = 0 otherwise. The initial configuration sets all $\mu_j = \mu_d <\mu_c$ yielding four visible MCMs labeled $\gamma_a$, $\gamma_b$, $\gamma_c$, $\gamma_d$, as shown in Fig. \ref{fig-5}(a).

To exchange $\gamma_a$ and $\gamma_b$, we adiabatically tune $\mu_2$, $\mu_3$, $\mu_4$ according to the sequence in Fig. \ref{fig-5}(i), while keeping the remaining $\mu_j$ fixed. This protocol moves the MCMs along the diagonals of $T_2$, $T_3$, $T_4$ and exchanges their positions [Figs. \ref{fig-5}(a–g)]. By the end of the cycle, the system parameters return to their initial values, but  $\gamma_a$ and $\gamma_b$ are exchanged.
Throughout the entire process, all four MCMs remain pinned at zero energy and are separated from excited states by a finite gap [red bands in Fig. \ref{fig-5}(h)]. The minimum gap along the protocol sets the adiabatic time scale, $T_{\mathrm{ad}} \gg \hbar / \Delta_{\min}$, which is readily compatible with gate-controlled dynamics in mesoscopic superconducting devices\cite{TS82, TS83, TS84, aasen2016milestones}. The same procedure applies to other configurations involving both active and inactive MCMs, enabling multi-MCM braiding through phase-controlled inter-island couplings (See SMs for details).

The stability of the zero modes against moderate variations of chemical potential and phase differences demonstrates that the operation is protected by the higher-order topological gap. In contrast to nanowire networks that require complex junctions, the present two-dimensional architecture supports dense packing of triangular islands and purely electrical control lines, offering a scalable route toward programmable Majorana networks.

\textit{Discussions and Conclusions.}--- Our proposal establishes AFMTIs as a promising materials platform for HOTS. The essential ingredient is the symmetry-enforced boundary dichotomy, which converts the coexistence of superconducting and magnetic masses into a source of MCMs. The approach is not restricted to MnXPb$_2$ but applies generically to AFMTIs with effective time-reversal symmetry and sizable bulk gaps. Experimentally, the proposed platform is compatible with existing thin-film growth and gating technologies. Proximity-induced gaps of order 0.5-1 meV and magnetic exchange fields several times larger ensure a robust topological regime accessible at dilution refrigerator temperatures. The electrical tunability of Majorana positions offers a direct route toward fusion and braiding experiments without magnetic fields or vortices.

In summary, we have introduced a symmetry-enforced route to HOTS based on AFMTIs proximitized by conventional superconductors, specifically, MnXPb$_2$ (X = Se, Te)-Pb heterostructure. The intrinsic boundary dichotomy of these systems—gapless Dirac states on antiferromagnetic edges and magnetic gaps on ferromagnetic edges—naturally generates MCMs as mass domain walls when superconductivity is induced. Using first-principles calculations and a calibrated effective theory, we demonstrated robust corner localization and purely electrical control of Majorana fusion and braiding in a triangular geometry.
Our results combine realistic materials, symmetry-protected boundary physics, and scalable electrical control, providing an experimentally accessible pathway toward programmable Majorana networks in two dimensions. More broadly, they establish AFMTIs as a general platform for HOTSs and open new opportunities for integrating magnetism, topology, and superconductivity in quantum devices.

\begin{acknowledgments}
This work was financially supported by the National Key R\&D Program of China (Grants No. 2024YFA1613200, No. 2022YFA1403200 and No. 2022YFA1403100), National Natural Science Foundation of
China (Grants No. 92265104, No. 12022413, No. 12175023), the Basic Research Program of the Chinese
Academy of Sciences Based on Major Scientific Infrastructures (Grant No. JZHKYPT-2021-08), the CASHIPS Director’s Fund (Grant No. BJPY2023A09), Anhui Provincial Major S\&T Project(s202305a12020005), and
the High Magnetic Field Laboratory of Anhui Province under Contract No. AHHM-FX-2020-02. 
\end{acknowledgments}

\bibliography{ref}

@article{PhysRevLett.114.017001,
  title = {Experimental Detection of a Majorana Mode in the core of a Magnetic Vortex inside a Topological Insulator-Superconductor ${\mathrm{Bi}}_{2}{\mathrm{Te}}_{3}/{\mathrm{NbSe}}_{2}$ Heterostructure},
  author = {Xu, Jin-Peng and Wang, Mei-Xiao and Liu, Zhi Long and Ge, Jian-Feng and Yang, Xiaojun and Liu, Canhua and Xu, Zhu An and Guan, Dandan and Gao, Chun Lei and Qian, Dong and Liu, Ying and Wang, Qiang-Hua and Zhang, Fu-Chun and Xue, Qi-Kun and Jia, Jin-Feng},
  journal = {Phys. Rev. Lett.},
  volume = {114},
  issue = {1},
  pages = {017001},
  numpages = {5},
  year = {2015},
  month = {Jan},
  publisher = {American Physical Society},
  doi = {10.1103/PhysRevLett.114.017001},
  url = {https://link.aps.org/doi/10.1103/PhysRevLett.114.017001}
}

@article{PhysRevLett.116.257003,
  title = {Majorana Zero Mode Detected with Spin Selective Andreev Reflection in the Vortex of a Topological Superconductor},
  author = {Sun, Hao-Hua and Zhang, Kai-Wen and Hu, Lun-Hui and Li, Chuang and Wang, Guan-Yong and Ma, Hai-Yang and Xu, Zhu-An and Gao, Chun-Lei and Guan, Dan-Dan and Li, Yao-Yi and Liu, Canhua and Qian, Dong and Zhou, Yi and Fu, Liang and Li, Shao-Chun and Zhang, Fu-Chun and Jia, Jin-Feng},
  journal = {Phys. Rev. Lett.},
  volume = {116},
  issue = {25},
  pages = {257003},
  numpages = {5},
  year = {2016},
  month = {Jun},
  publisher = {American Physical Society},
  doi = {10.1103/PhysRevLett.116.257003},
  url = {https://link.aps.org/doi/10.1103/PhysRevLett.116.257003}
}

@article{sau2010generic,
  title={Generic New Platform for Topological Quantum Computation Using Semiconductor Heterostructures},
  author={Sau, Jay D and Lutchyn, Roman M and Tewari, Sumanta and Das Sarma, Sankar},
  journal={Physical review letters},
  volume={104},
  number={4},
  pages={040502},
  year={2010},
  publisher={APS}
}

@article{nadj2014observation,
  title={Observation of Majorana fermions in ferromagnetic atomic chains on a superconductor},
  author={Nadj-Perge, Stevan and Drozdov, Ilya K and Li, Jian and Chen, Hua and Jeon, Sangjun and Seo, Jungpil and MacDonald, Allan H and Bernevig, B Andrei and Yazdani, Ali},
  journal={Science},
  volume={346},
  number={6209},
  pages={602--607},
  year={2014},
  publisher={American Association for the Advancement of Science}
}

@article{hao2014topological,
  title={Topological phases in the single-layer FeSe},
  author={Hao, Ningning and Hu, Jiangping},
  journal={Physical Review X},
  volume={4},
  number={3},
  pages={031053},
  year={2014},
  publisher={APS}
}

@article{wang2015topological,
  title={Topological nature of the FeSe 0.5 Te 0.5 superconductor},
  author={Wang, Zhijun and Zhang, Peng and Xu, Gang and Zeng, LK and Miao, Hu and Xu, Xiaoyan and Qian, Tian and Weng, Hongming and Richard, P and Fedorov, Alexei V and others},
  journal={Physical Review B},
  volume={92},
  number={11},
  pages={115119},
  year={2015},
  publisher={APS}
}

@article{wu2016topological,
  title={Topological characters in Fe (Te 1- x Se x) thin films},
  author={Wu, Xianxin and Qin, Shengshan and Liang, Yi and Fan, Heng and Hu, Jiangping},
  journal={Physical Review B},
  volume={93},
  number={11},
  pages={115129},
  year={2016},
  publisher={APS}
}

@article{hao2019topological,
  title={Topological quantum states of matter in iron-based superconductors: from concept to material realization},
  author={Hao, Ning and Hu, Jiangping},
  journal={National Science Review},
  volume={6},
  number={2},
  pages={213--226},
  year={2019},
  publisher={Oxford University Press}
}

@article{wu2015cafeas,
  title={CaFeAs 2: A staggered intercalation of quantum spin Hall and high-temperature superconductivity},
  author={Wu, Xianxin and Qin, Shengshan and Liang, Yi and Le, Congcong and Fan, Heng and Hu, Jiangping},
  journal={Physical Review B},
  volume={91},
  number={8},
  pages={081111},
  year={2015},
  publisher={APS}
}

@article{xu2016topological,
  title={Topological superconductivity on the surface of Fe-based superconductors},
  author={Xu, Gang and Lian, Biao and Tang, Peizhe and Qi, Xiao-Liang and Zhang, Shou-Cheng},
  journal={Physical review letters},
  volume={117},
  number={4},
  pages={047001},
  year={2016},
  publisher={APS}
}

@article{karzig2017scalable,
  title={Scalable designs for quasiparticle-poisoning-protected topological quantum computation with Majorana zero modes},
  author={Karzig, Torsten and Knapp, Christina and Lutchyn, Roman M and Bonderson, Parsa and Hastings, Matthew B and Nayak, Chetan and Alicea, Jason and Flensberg, Karsten and Plugge, Stephan and Oreg, Yuval and others},
  journal={Physical Review B},
  volume={95},
  number={23},
  pages={235305},
  year={2017},
  publisher={APS}
}

@article{PhysRevLett.124.227001,
  title = {In-Plane Zeeman-Field-Induced Majorana Corner and Hinge Modes in an $s$-Wave Superconductor Heterostructure},
  author = {Wu, Ya-Jie and Hou, Junpeng and Li, Yun-Mei and Luo, Xi-Wang and Shi, Xiaoyan and Zhang, Chuanwei},
  journal = {Phys. Rev. Lett.},
  volume = {124},
  issue = {22},
  pages = {227001},
  numpages = {7},
  year = {2020},
  month = {Jun},
  publisher = {American Physical Society},
  doi = {10.1103/PhysRevLett.124.227001},
  url = {https://link.aps.org/doi/10.1103/PhysRevLett.124.227001}
}

@article{PhysRevLett.123.167001,
  title = {Higher-Order Topology and Nodal Topological Superconductivity in Fe(Se,Te) Heterostructures},
  author = {Zhang, Rui-Xing and Cole, William S. and Wu, Xianxin and Das Sarma, S.},
  journal = {Phys. Rev. Lett.},
  volume = {123},
  issue = {16},
  pages = {167001},
  numpages = {6},
  year = {2019},
  month = {Oct},
  publisher = {American Physical Society},
  doi = {10.1103/PhysRevLett.123.167001},
  url = {https://link.aps.org/doi/10.1103/PhysRevLett.123.167001}
}

@ARTICLE{TS1,
	author={Lutchyn, R. M. and Bakkers, E. P. A. M. and Kouwenhoven, L. P. and Krogstrup, P. and Marcus, C. M. and Oreg, Y.},
	title={Majorana zero modes in superconductor–semiconductor heterostructures},
	journal={ Nat. Rev. Mater.},
	year = {2018},
	volume = {3},
	pages = {52--68},
	number = {5},
	doi = {10.1038/s41578-018-0003-1},
}

@article{TS2,
	title = {Non-Abelian anyons and topological quantum computation},
	author = {Nayak, Chetan and Simon, Steven H. and Stern, Ady and Freedman, Michael and Das Sarma, Sankar},
	journal = {Rev. Mod. Phys.},
	volume = {80},
	issue = {3},
	pages = {1083--1159},
	numpages = {0},
	year = {2008},
	month = {Sep},
	publisher = {American Physical Society},
	doi = {10.1103/RevModPhys.80.1083},
}

@article{TS3,
	doi = {10.1088/1361-6633/aa6ac7},
	url = {https://dx.doi.org/10.1088/1361-6633/aa6ac7},
	year = {2017},
	month = {may},
	publisher = {IOP Publishing},
	volume = {80},
	number = {7},
	pages = {076501},
	author = {Masatoshi Sato and Yoichi Ando},
	title = {Topological superconductors: a review},
	journal = {Rep. Prog. Phys.},
}

@article{TS4,
	doi = {10.1070/1063-7869/44/10S/S29},
	url = {https://dx.doi.org/10.1070/1063-7869/44/10S/S29},
	year = {2001},
	month = {oct},
	publisher = {},
	volume = {44},
	number = {10S},
	pages = {131},
	author = {A Yu Kitaev},
	title = {Unpaired Majorana fermions in quantum wires},
	journal = {Phys. Uspekhi},

}

@article{TS5,
	doi = {10.1088/1367-2630/14/3/035019},
	url = {https://dx.doi.org/10.1088/1367-2630/14/3/035019},
	year = {2012},
	month = {mar},
	publisher = {IOP Publishing},
	volume = {14},
	number = {3},
	pages = {035019},
	author = {B van Heck and A R Akhmerov and F Hassler and M Burrello and C W J Beenakker},
	title = {Coulomb-assisted braiding of Majorana fermions in a Josephson junction array},
	journal = {New J. Phys.},
}

@article{TS6,
	title = {Flux-controlled quantum computation with Majorana fermions},
	author = {Hyart, T. and van Heck, B. and Fulga, I. C. and Burrello, M. and Akhmerov, A. R. and Beenakker, C. W. J.},
	journal = {Phys. Rev. B},
	volume = {88},
	issue = {3},
	pages = {035121},
	numpages = {17},
	year = {2013},
	month = {Jul},
	publisher = {American Physical Society},
	doi = {10.1103/PhysRevB.88.035121},
}

@article{TS7,
	title = {Towards Realistic Implementations of a Majorana Surface Code},
	author = {Landau, L. A. and Plugge, S. and Sela, E. and Altland, A. and Albrecht, S. M. and Egger, R.},
	journal = {Phys. Rev. Lett.},
	volume = {116},
	issue = {5},
	pages = {050501},
	numpages = {5},
	year = {2016},
	month = {Feb},
	publisher = {American Physical Society},
	doi = {10.1103/PhysRevLett.116.050501},
}

@article{TS8,
	title = {Milestones Toward Majorana-Based Quantum Computing},
	author = {Aasen, David and Hell, Michael and Mishmash, Ryan V. and Higginbotham, Andrew and Danon, Jeroen and Leijnse, Martin and Jespersen, Thomas S. and Folk, Joshua A. and Marcus, Charles M. and Flensberg, Karsten and Alicea, Jason},
	journal = {Phys. Rev. X},
	volume = {6},
	issue = {3},
	pages = {031016},
	numpages = {28},
	year = {2016},
	month = {Aug},
	publisher = {American Physical Society},
	doi = {10.1103/PhysRevX.6.031016},
}

@article{TS9,
	title = {Fault-tolerant quantum computation by anyons},
	journal = {Ann. Phys.},
	volume = {303},
	number = {1},
	pages = {2-30},
	year = {2003},
	issn = {0003-4916},
	doi = {https://doi.org/10.1016/S0003-4916(02)00018-0},
	url = {https://www.sciencedirect.com/science/article/pii/S0003491602000180},
	author = {A.Yu. Kitaev},
}

@article{TS10,
	title = {Topological insulators and superconductors},
	author = {Qi, Xiao-Liang and Zhang, Shou-Cheng},
	journal = {Rev. Mod. Phys.},
	volume = {83},
	issue = {4},
	pages = {1057--1110},
	numpages = {0},
	year = {2011},
	month = {Oct},
	publisher = {American Physical Society},
	doi = {10.1103/RevModPhys.83.1057},
	url = {https://link.aps.org/doi/10.1103/RevModPhys.83.1057}
}

@article{TS35,
	title = {Superconducting Proximity Effect and Majorana Fermions at the Surface of a Topological Insulator},
	author = {Fu, Liang and Kane, C. L.},
	journal = {Phys. Rev. Lett.},
	volume = {100},
	issue = {9},
	pages = {096407},
	numpages = {4},
	year = {2008},
	month = {Mar},
	publisher = {American Physical Society},
	doi = {10.1103/PhysRevLett.100.096407},
	url = {https://link.aps.org/doi/10.1103/PhysRevLett.100.096407}
}

@article{TS75,
	author = {B. Andrei Bernevig  and Taylor L. Hughes  and Shou-Cheng Zhang },
	title = {Quantum Spin Hall Effect and Topological Phase Transition in HgTe Quantum Wells},
	journal = {Science},
	volume = {314},
	number = {5806},
	pages = {1757-1761},
	year = {2006},
	doi = {10.1126/science.1133734},
	URL = {https://www.science.org/doi/abs/10.1126/science.1133734},
	eprint = {https://www.science.org/doi/pdf/10.1126/science.1133734},
}

@article{TS82,
	title = {All-electrically tunable networks of Majorana bound states},
	author = {Zhang, Song-Bo and Calzona, Alessio and Trauzettel, Bj\"orn},
	journal = {Phys. Rev. B},
	volume = {102},
	issue = {10},
	pages = {100503},
	numpages = {7},
	year = {2020},
	month = {Sep},
	publisher = {American Physical Society},
	doi = {10.1103/PhysRevB.102.100503},
	url = {https://link.aps.org/doi/10.1103/PhysRevB.102.100503}
}

@article{TS83,
	title = {Topological and holonomic quantum computation based on second-order topological superconductors},
	author = {Zhang, Song-Bo and Rui, W. B. and Calzona, Alessio and Choi, Sang-Jun and Schnyder, Andreas P. and Trauzettel, Bj\"orn},
	journal = {Phys. Rev. Res.},
	volume = {2},
	issue = {4},
	pages = {043025},
	numpages = {17},
	year = {2020},
	month = {Oct},
	publisher = {American Physical Society},
	doi = {10.1103/PhysRevResearch.2.043025},
	url = {https://link.aps.org/doi/10.1103/PhysRevResearch.2.043025}
}

@article{TS84,
	title = {Non-Abelian statistics and topological quantum information processing in 1D wire networks},
	author = {Alicea, Jason and Oreg, Yuval and Refael, Gil and von Oppen, Felix and Fisher, Matthew P. A.},
	journal = {Nat. Phys.},
	volume = {7},
	issue = {5},
	pages = {412-417},
	year = {2011},
	month = {May},
	doi = {10.1038/nphys1915},
	url = {https://doi.org/10.1038/nphys1915}
}

@Article{TS85,
	title={{Search for non-Abelian Majorana braiding statistics in superconductors}},
	author={C. W. J. Beenakker},
	journal={SciPost Phys. Lect. Notes},
	doi={10.21468/SciPostPhysLectNotes.15},
	url={https://scipost.org/10.21468/SciPostPhys
	pages={15},
	year={2020},
	publisher={SciPost},LectNotes.15},
}

@article{TS86,
	title = {Majorana fermion exchange in quasi-one-dimensional networks},
	author = {Clarke, David J. and Sau, Jay D. and Tewari, Sumanta},
	journal = {Phys. Rev. B},
	volume = {84},
	issue = {3},
	pages = {035120},
	numpages = {8},
	year = {2011},
	month = {Jul},
	publisher = {American Physical Society},
	doi = {10.1103/PhysRevB.84.035120},
	url = {https://link.aps.org/doi/10.1103/PhysRevB.84.035120}
}

@article{TS87,
	title = {Majorana braiding in realistic nanowire Y-junctions and tuning forks},
	author = {Harper, Fenner and Pushp, Aakash and Roy, Rahul},
	journal = {Phys. Rev. Res.},
	volume = {1},
	issue = {3},
	pages = {033207},
	numpages = {16},
	year = {2019},
	month = {Dec},
	publisher = {American Physical Society},
	doi = {10.1103/PhysRevResearch.1.033207},
	url = {https://link.aps.org/doi/10.1103/PhysRevResearch.1.033207}
}

@article{TS88,
	title = {Manipulating and braiding Majorana corner modes on a kagome lattice},
	author = {He, San-Ren and Bi, Xiu-Lian and Fan, Jing and Wang, Zhen-Hua and Li, Lin and Xu, Dong-Hui},
	journal = {Phys. Rev. B},
	volume = {111},
	issue = {14},
	pages = {144506},
	numpages = {10},
	year = {2025},
	month = {Apr},
	publisher = {American Physical Society},
	doi = {10.1103/PhysRevB.111.144506},
	url = {https://link.aps.org/doi/10.1103/PhysRevB.111.144506}
}

@article{TS89,
	title = {Interplay between the atomic structures and superconductivity of two-monolayer Pb films},
	author = {Xie, Kun and Huang, Ze and Li, Pengju and Xia, Yumin and Cai, Desheng and Gu, Yitong and Liu, Yuzhou and Cai, Fangliang and Zhang, Runxiao and Shi, Haohao and Cui, Ping and Qin, Shengyong},
	journal = {Sci. China- Phys. Mech. Astron},
	volume = {67},
	issue = {1},
	pages = {217411},
	numpages = {1869-1927},
	year = {2023},
	month = {Dec},
	doi = {10.1007/s11433-023-2191-5},
	url = {https://doi.org/10.1007/s11433-023-2191-5}
}

@article{PhysRevB.98.165144,
  title = {Weak-pairing higher order topological superconductors},
  author = {Wang, Yuxuan and Lin, Mao and Hughes, Taylor L.},
  journal = {Phys. Rev. B},
  volume = {98},
  issue = {16},
  pages = {165144},
  numpages = {13},
  year = {2018},
  month = {Oct},
  publisher = {American Physical Society},
  doi = {10.1103/PhysRevB.98.165144},
  url = {https://link.aps.org/doi/10.1103/PhysRevB.98.165144}
}

@article{wang2018evidence,
  title={Evidence for Majorana bound states in an iron-based superconductor},
  author={Wang, Dongfei and Kong, Lingyuan and Fan, Peng and Chen, Hui and Zhu, Shiyu and Liu, Wenyao and Cao, Lu and Sun, Yujie and Du, Shixuan and Schneeloch, John and others},
  journal={Science},
  volume={362},
  number={6412},
  pages={333--335},
  year={2018},
  publisher={American Association for the Advancement of Science}
}

@article{song2022phase,
  title={Phase-manipulation-induced Majorana mode and braiding realization in iron-based superconductor Fe (Te, Se)},
  author={Song, Rui and Zhang, Ping and Hao, Ning},
  journal={Physical Review Letters},
  volume={128},
  number={1},
  pages={016402},
  year={2022},
  publisher={APS}
}

@article{li2022ordered,
  title={Ordered and tunable Majorana-zero-mode lattice in naturally strained LiFeAs},
  author={Li, Meng and Li, Geng and Cao, Lu and Zhou, Xingtai and Wang, Xiancheng and Jin, Changqing and Chiu, Ching-Kai and Pennycook, Stephen J and Wang, Ziqiang and Gao, Hong-Jun},
  journal={Nature},
  volume={606},
  number={7916},
  pages={890--895},
  year={2022},
  publisher={Nature Publishing Group UK London}
}

@article{liu2024signatures,
  title={Signatures of hybridization of multiple Majorana zero modes in a vortex},
  author={Liu, Tengteng and Wan, Chun Yu and Yang, Hao and Zhao, Yujun and Xie, Bangjin and Zheng, Weiyan and Yi, Zhaoxia and Guan, Dandan and Wang, Shiyong and Zheng, Hao and others},
  journal={Nature},
  volume={633},
  number={8028},
  pages={71--76},
  year={2024},
  publisher={Nature Publishing Group UK London}
}

@article{yan2019higher,
  title={Higher-order topological odd-parity superconductors},
  author={Yan, Zhongbo},
  journal={Physical review letters},
  volume={123},
  number={17},
  pages={177001},
  year={2019},
  publisher={APS}
}

@article{zhang2020higher,
  title={Higher-order topological Dirac superconductors},
  author={Zhang, Rui-Xing and Hsu, Yi-Ting and Das Sarma, S},
  journal={Physical Review B},
  volume={102},
  number={9},
  pages={094503},
  year={2020},
  publisher={APS}
}

@article{khalaf2018higher,
  title={Higher-order topological insulators and superconductors protected by inversion symmetry},
  author={Khalaf, Eslam},
  journal={Physical Review B},
  volume={97},
  number={20},
  pages={205136},
  year={2018},
  publisher={APS}
}

@article{scammell2022intrinsic,
  title={Intrinsic first-and higher-order topological superconductivity in a doped topological insulator},
  author={Scammell, Harley D and Ingham, Julian and Geier, Max and Li, Tommy},
  journal={Physical Review B},
  volume={105},
  number={19},
  pages={195149},
  year={2022},
  publisher={APS}
}

@article{ahn2020higher,
  title={Higher-order topological superconductivity of spin-polarized fermions},
  author={Ahn, Junyeong and Yang, Bohm-Jung},
  journal={Physical Review Research},
  volume={2},
  number={1},
  pages={012060},
  year={2020},
  publisher={APS}
}

@article{li2024realizing,
  title={Realizing tunable higher-order topological superconductors with altermagnets},
  author={Li, Yu-Xuan},
  journal={Physical Review B},
  volume={109},
  number={22},
  pages={224502},
  year={2024},
  publisher={APS}
}

@article{wong2023higher,
  title={Higher order topological superconductivity in magnet-superconductor hybrid systems},
  author={Wong, Ka Ho and Hirsbrunner, Mark R and Gliozzi, Jacopo and Malik, Arbaz and Bradlyn, Barry and Hughes, Taylor L and Morr, Dirk K},
  journal={npj Quantum Materials},
  volume={8},
  number={1},
  pages={31},
  year={2023},
  publisher={Nature Publishing Group UK London}
}

@article{PhysRevB.108.184517,
  title = {Higher-order topological superconductor phases in a multilayer system},
  author = {Nakai, Ryota and Nomura, Kentaro},
  journal = {Phys. Rev. B},
  volume = {108},
  issue = {18},
  pages = {184517},
  numpages = {9},
  year = {2023},
  month = {Nov},
  publisher = {American Physical Society},
  doi = {10.1103/PhysRevB.108.184517},
  url = {https://link.aps.org/doi/10.1103/PhysRevB.108.184517}
}

@article{volpez2019,
  title={Second-order topological superconductivity in $\pi$-junction Rashba layers},
  author={Volpez, Yanick and Loss, Daniel and Klinovaja, Jelena},
  journal={Physical review letters},
  volume={122},
  number={12},
  pages={126402},
  year={2019},
  publisher={APS}
}

@article{wu2020boundary,
  title={Boundary-obstructed topological high-T c superconductivity in iron pnictides},
  author={Wu, Xianxin and Benalcazar, Wladimir A and Li, Yinxiang and Thomale, Ronny and Liu, Chao-Xing and Hu, Jiangping},
  journal={Physical Review X},
  volume={10},
  number={4},
  pages={041014},
  year={2020},
  publisher={APS}
}

@article{pan2019lattice,
  title={Lattice-symmetry-assisted second-order topological superconductors and Majorana patterns},
  author={Pan, Xiao-Hong and Yang, Kai-Jie and Chen, Li and Xu, Gang and Liu, Chao-Xing and Liu, Xin},
  journal={Physical review letters},
  volume={123},
  number={15},
  pages={156801},
  year={2019},
  publisher={APS}
}

@article{machida2019zero,
  title={Zero-energy vortex bound state in the superconducting topological surface state of Fe (Se, Te)},
  author={Machida, T and Sun, Y and Pyon, S and Takeda, S and Kohsaka, Y and Hanaguri, T and Sasagawa, T and Tamegai, T},
  journal={Nature materials},
  volume={18},
  number={8},
  pages={811--815},
  year={2019},
  publisher={Nature Publishing Group UK London}
}

@article{oyler2009chemical,
  title={Chemical synthesis of two-dimensional iron chalcogenide nanosheets: FeSe, FeTe, Fe (Se, Te), and FeTe2},
  author={Oyler, Karl D and Ke, Xianglin and Sines, Ian T and Schiffer, Peter and Schaak, Raymond E},
  journal={Chemistry of Materials},
  volume={21},
  number={15},
  pages={3655--3661},
  year={2009},
  publisher={ACS Publications}
}

@article{aasen2016milestones,
  title={Milestones toward Majorana-based quantum computing},
  author={Aasen, David and Hell, Michael and Mishmash, Ryan V and Higginbotham, Andrew and Danon, Jeroen and Leijnse, Martin and Jespersen, Thomas S and Folk, Joshua A and Marcus, Charles M and Flensberg, Karsten and others},
  journal={Physical Review X},
  volume={6},
  number={3},
  pages={031016},
  year={2016},
  publisher={APS}
}

\end{document}